Choosing an Optimal Method for Causal Decomposition Analysis:

A Better Practice for Identifying Contributing Factors to Health Disparities

Soojin Park[1]

Suyeon Kang

Chioun Lee

University of California-Riverside,

[1]Email: soojinp@ucr.edu, Phone: 951-827-1504

Abstract

Causal decomposition analysis provides a way to identify mediators that contribute to health disparities between marginalized and non-marginalized groups. In particular, the degree to which a disparity would be reduced/remain after intervening on a mediator is of interest. Yet, estimating disparity reduction/remaining might be challenging for many researchers, possibly because there is a lack of understanding of how each estimation method differs from other methods. In addition, there is no appropriate estimation method available for a certain setting (i.e., a regression-based approach with a categorical mediator). Therefore, we review the merits and limitations of the existing three estimation methods (i.e., regression, weighting, and imputation) and provide two new extensions that are useful in practical settings. A flexible new method uses an extended imputation approach to address a categorical and continuous mediator/outcome while incorporating any nonlinear relationships. A new regression method provides a simple estimator that performs well in terms of bias and variance but at the cost of assuming linearity, except for exposure–mediator interactions. Recommendations are given for choosing methods based on a review of different methods and simulation studies. We demonstrate the practice of choosing an optimal method by identifying mediators that reduce race–gender disparity in cardiovascular health, using data from the Midlife Development in the US study. We also offer open-source software for R (*causal.decomp*) that implements some estimation methods presented in the study.

Choosing an Optimal Method for Causal Decomposition Analysis:

A Better Practice for Identifying Contributing Factors to Health Disparities

## 1. Introduction

Across disciplines, for example, health psychology, medical sociology, and epidemiology, researchers have been investigating the degree to which health disparities exist between marginalized and non-marginalized groups, as defined by social characteristics, such as gender, race/ethnicity, socioeconomic status (SES), and sexual orientation. More importantly, health disparity researchers across disciplines have been interested in identifying mediators that produce disparities. One statistical framework that allows researchers to identify mediators underlying disparities is causal decomposition analysis (Jackson & VanderWeele, 2018). Causal decomposition analysis was developed under the potential outcomes framework (Rubin, 1978), which facilitates deriving a valid conclusion by precisely defining causal estimands (i.e., effects of interest), identifying assumptions that allow a causal interpretation of the estimands, and providing a flexible estimation method that allows nonlinear relationships. In addition, causal decomposition analysis approaches mediation from an interventional perspective—i.e., how intervening on mediators in a certain way would change disparities (Nguyen, Schmid, & Stuart, 2020). In line with this interventional perspective, *disparity reduction/remaining* is defined as the degree to which the disparity would be reduced or remain if we hypothetically intervened to make the distribution of mediators equal between marginalized and non-marginalized groups. This intervention, although hypothetical, is policy-relevant since it requires adjusting the level of mediators (such as college education) of a marginalized group to the level of a non-marginalized group.

Recently, many estimation methods have been developed for estimating this disparity reduction/remaining. The developed methods include a regression-based approach (Jackson & VanderWeele, 2018), a weighting-based approach (Jackson, 2019), and an imputation-based approach (Park, Lee, & Qin, 2020) [1]. Yet, estimating disparity reduction/remaining might be challenging for many researchers possibly because there is 1) a lack of understanding of how each estimation method differs and performs compared to other methods and 2) a lack of estimation methods in certain settings (e.g., a regression-based approach with a categorical mediator). While all three approaches can reasonably estimate disparity reduction/remaining, they vary in their advantages and disadvantages, depending on the data at hand and the researchers' assumptions. Therefore, we discuss each method's merits and limitations based on a review of the methods and small-scale simulation studies. We also provide two extensions to existing methods that are useful in settings different from those in the aforementioned studies. As a result, the current study aims to offer general guidelines for choosing an optimal method to estimate disparity reduction/remaining.

In the following section (Section 2), we will present our motivating example. In Section 3, we will briefly introduce causal decomposition analysis in the context of the motivating example. In Section 4, we will review three estimation methods that can be used to quantify disparity reduction/remaining and present two extensions to the existing methods that are useful. In Section 5, we will conduct small-scale simulation studies to compare performances of the five methods. Based on a review of the methods and small-scale simulation studies, we provide recommendations for selecting optimal estimation methods in the context of the motivation example (Section 6). In Section 7, we conclude with a discussion.

Open-source software for R (*causal.decomp*) that implements some of the estimation methods presented here is available from XX [to be updated]. Code to replicate all analyses can be found in Supplementary Materials.

---

[1] Previously, Park and her colleagues (Park et al., 2020) developed an imputation and weighting method, but here we call it an imputation method since weighting is not used for the effects of interest (i.e., conditional disparity reduction/remaining) in this study.

## 2. Motivating Example

As a motivating example, we investigate gendered racial disparities in cardiovascular health (CVH). Despite robust declines in cardiovascular disease (CVD) mortality in the US over 50 years, racial and ethnic differences in the burden of CVD are still prominent. To improve CVD free longevity and reduce health disparities in populations, the American Heart Association (AHA) introduced the ideal CVH metric (known as Life's Simple 7) in 2010 to encourage lifestyle changes, including blood pressure management, cholesterol control, blood glucose reduction, physical activity, healthy diet, weight loss, and tobacco cessation. A decade of evidence, however, shows that Blacks, particularly Black women, are far less likely than Whites to achieve ideal CVH scores. While observing CVH disparities between Black women and Whites is important, Lee, Park, and Boylan (2021) seek to investigate mediators that produce such disparities.

Prior studies have found that socioeconomic status (SES) and discrimination partially explain racial/ethnic disparities in CVH (Bey, Jesdale, Forrester, Person, & Kiefe, 2019). There has also been growing interest in the role of early-life environments and circumstances (Suglia et al., 2018). Specifically, throughout the life course, Blacks are socially disadvantaged relative to Whites in the US across multiple domains, including educational attainment, income, and wealth (Maxwell, 1994; Oliver & Shapiro, 2013; Pollack et al., 2013), as well as health care access and insurance coverage (Lê Cook, McGuire, & Zuvekas, 2009; Sohn, 2017). Women of color (e.g., Black women) are particularly likely to experience insecure economic positions (Brown, 2012). However, racial/ethnic differences in health outcomes persist even after accounting for SES (Hayward, Miles, Crimmins, & Yang, 2000; Williams, Priest, & Anderson, 2019), and a growing body of literature supports the notion that exposure to interpersonal racial discrimination contributes to health disparities (Williams & Sternthal, 2010). Moreover, some stressful events in early life, including sexual abuse, are more common for racial or gender minorities. Such trauma might negatively influence cognitive and socioemotional development and increase the likelihood of engaging in unhealthy behaviors and lifestyles (Miller, Chen, & Parker, 2011; Shonkoff, Boyce, & McEwen, 2009), which may in turn have life-long consequences for CVH.

Lee et al. (2021) used a sample of 1978 respondents from the Midlife Development in the US (MIDUS) study and the MIDUS Refresher who identified themselves either as non-Hispanic White or non-Hispanic Black. Racial and gender statuses were created using the nexus of self-identified race/ethnicity and gender (i.e., White men, White women, Black men, and Black women). CVH was assessed in accordance with the AHA's criteria for seven components: smoking, BMI, physical activity, diet, total cholesterol, blood pressure, and fasting glucose. A composite score of CVH was created where higher values indicate better CVH that sums the criteria for ideal, intermediate, or poor CVH for each component.

To test the mechanisms, the authors considered three life-course mediators (childhood abuse, perceived discrimination, and education) that explain cardiovascular disparities across gender and racial groups. *Childhood abuse* is an index measuring experiences of emotional, physical, or sexual abuse, with possible responses to each item ranging from 1 (never true) to 5 (very often true). As for *perceived discrimination*, respondents were asked to report the number of times in their life they faced "discrimination" in each of 11 domains. *Education* is a variable that indicates the highest level of degree completed, which ranges from 1 = no school/some grade school to 12 = PhD, MD, or other professional degree. We use this motivating example throughout the manuscript.

Figure 1 shows a Directed Acyclic Graph (DAG) in which nodes represent variables and arrows represent causal effects. It is important to delineate our hypothesized causal structure between variables because we formulate research questions and determine variables that should be controlled for based on this hypothesized causal structure. The race-gender variable is denoted as $R$; CVH is denoted as $Y$. On the pathway from race-gender ($R$) to CVH ($Y$), we hypothesize three mediators: childhood abuse ($M_1$),

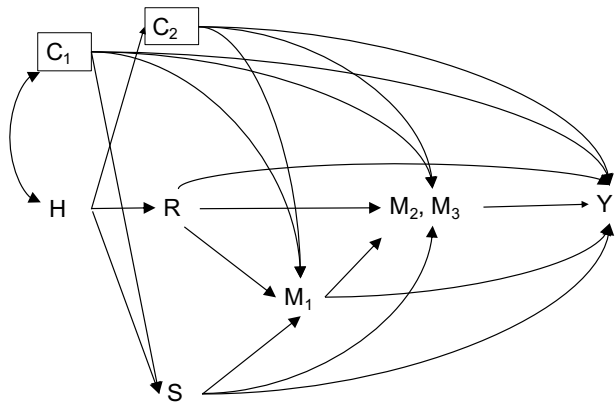

*Figure 1*. Directed acyclic graph showing the relationship between intersectional status, cardiovascular health, and three potential mediators

Note. 1) Diagram represents the relationship between race and gender intersectional status $R$, cardiovascular health $Y$, child abuse $M_1$, perceived discrimination $M_2$, and education $M_3$, as well as history $H$, age $C_1$, genetic vulnerability $C_2$, and childhood SES $S$. 2) Placing a box around the conditioning variables implies that a disparity is considered within levels of these variables.

discrimination ($M_2$), and education ($M_3$). In line with Kaufman (2008), we assume that childhood socioeconomic status ($S$) is correlated with race through historical processes ($H$) that include racism. We also assume that baseline covariates (age $C_1$ and genetic vulnerability $C_2$) are correlated with the historical processes. For simplicity, we use $C$ to encompass both $C_1$ and $C_2$.

Note that there is no arrow between discrimination ($M_2$) and education ($M_3$). Unlike the previous study, we do not specify a causal ordering between these mediators because the relationship between the mediators could be bidirectional. For example, one could imagine that poor education could exacerbate discrimination. On the contrary, discrimination could lead to lower levels of educational achievement (Park et al., 2020). One way to address this unclear or bidirectional causal ordering of mediators is simultaneously intervening on these multiple mediators (VanderWeele & Vansteelandt, 2014).

Given this DAG, we formulated the following research questions: 1) to what extent would equalizing the exposure to childhood abuse across race-gender groups reduce CVH disparity, and 2) to what extent equalizing education and perceived discrimination simultaneously across race-gender groups reduce CVH disparity? In line with these questions, we consider two hypothetical interventions of equalizing distributions across groups of the following mediators: 1) *childhood abuse* and 2) *education and discrimination.* For each intervention, we controlled for the same baseline covariates (i.e., age and genetic vulnerability) but different intermediate confounders. Intermediate confounders are the variables that confound the mediator-outcome relationship and are measured after or concurrently with the exposure (race or gender). For intervention 1, childhood SES ($S$) serves as an intermediate confounder, and for intervention 2, childhood SES ($S$) and childhood abuse ($M_1$) serve as intermediate confounders.

## 3. Causal Decomposition Analysis: Review

In this section, we first begin by quantifying the observed disparity between marginalized and non-marginalized groups. Then, we move on to decomposing the observed disparity into disparity reduction and remaining due to a hypothetical intervention. Throughout the paper, we focus on disparities conditional on baseline covariates to ensure comparability across different estimation methods. However, marginal disparities can be obtained by averaging over baseline covariates.

**Initial Disparity.** The observed disparity is defined as the average difference in an outcome between marginalized and non-marginalized groups among those who have the same value of baseline covariates. Suppose that we are interested in the CVH disparity between Black women ($R = 1$; comparison group) and White men ($R = 0$; reference group) among those who have the same age and genetic vulnerability. Then, the initial disparity for Black women compared to White men is formally defined as $\tau(1, 0) \equiv E[Y|R = 1, c] - E[Y|R = 0, c]$, where $c \in \mathcal{C}$. Note that the defined initial disparity is simply the observed mean difference in an outcome between marginalized and non-marginalized groups given baseline covariates. Thus, this is not causal unless assumptions are invoked, such as no omitted confounding. While the causal effect of race or gender is certainly conceivable, we do not attempt to measure this causal effect of race or gender since social-demographic factors are essentially non-modifiable (VanderWeele & Robinson, 2014).

**Intervention 1.** Once we observed the disparity between Black women and White men, we would also want to know how to reduce the disparity, for example, by reducing Black women's exposure to childhood abuse to the level of White men. Statistically, this hypothetical intervention requires assigning Black women's abuse with values randomly drawn from the distribution of White men's abuse. As shown in Table 1, the disparity reduction due to intervening on childhood abuse contrasts the following two conditions: 1) CVH for Black women and 2) CVH for Black women after assigning their abuse with random draws from the distribution of White men's abuse. The difference between the two conditions (1-2) estimates how much the disparity for Black women would be reduced if we hypothetically intervene to decrease Black women's abuse to the level of White men. Likewise, the disparity remaining contrasts the following two conditions: 2) CVH for Black women after assigning their abuse with random draws from the distribution of White men's abuse and 3) CVH for White men. The difference between the two conditions (2-3) estimates how much the disparity for Black women would remain even after the hypothetical intervention to decrease Black women's abuse to the level of White men.

Table 1
*Decomposition of the initial disparity*

|   | Observed and Potential Outcomes | Notation |
|---|---|---|
| 1 | Expected CVH for Black women | $E[Y\|R=1,c]$ |
| 2 | Expected CVH for Black women after decreasing their abuse to the level of White men | $E[Y(G_{m_1\|c}(0))\|R=1,c]$ |
| 3 | Expected CVH for White men | $E[Y\|R=0,c]$ |

1) Initial disparity $\tau(1,0)$= 1-3, Disparity reduction $\delta^1(1)$= 1-2, and Disparity remaining $\zeta^1(0)$ =2-3.
2) Observed and potential outcomes are assumed to be conditional on covariates.

Formally, let a random value drawn from the distribution of White men's abuse given baseline covariates be denoted as $G_{m_1|c}(0)$. Then, we can denote a potential CVH outcome for Black women if their abuse level was the same as White men who have the same age and genetic vulnerability as $Y(G_{m_1|c}(0)|R = 1, c)$. Disparity reduction and remaining are formally defined as $\delta^1(1) \equiv E[Y|R = 1, c] - E[Y(G_{m_1|c}(0))|R = 1, c]$ and $\zeta^1(0) \equiv E[Y(G_{m_1|c}(0))|R = 1, c] - E[Y|R = 0, c]$, respectively. The initial disparity can be obtained by summing disparity reduction and remaining as $\tau(1, 0) = \delta^1(1) + \zeta^1(0)$.

This definition of disparity reduction/remaining involves an unobservable potential outcome–i.e., $Y(G_{m_1|c}(0))$. Therefore, we need to invoke assumptions to convert this potential outcome to an observable quantity. The assumptions include 1) conditional independence, 2) positivity, and 3) consistency. The conditional independence states no omitted confounding in the mediator-outcome relationship given race-gender status $(R)$, intermediate confounder $(S)$, and baseline covariates $(C)$. The positivity assumes 1) a positive conditional probability of race-gender status given covariates and 2) a positive conditional probability among Black women $(R = 1)$ of each observed value for the mediator given covariates. Finally, the consistency assumes that the observed outcome under a particular exposure value is the same as the outcome after intervening to set the exposure to that value. All these assumptions are strong and whether the assumptions are met or not depends on a substantive example. The identification results are shown in Supplementary Materials.

**Intervention 2.** The disparity reduction/remaining due to multiple mediators (i.e., education and discrimination) can be defined the same way. Disparity reduction/remaining are defined as how much the disparity would be reduced/remain if we hypothetically intervene to equalize education and discrimination simultaneously between Black women and White men. Formal definitions as well as identification assumptions and results for intervention 2 are shown in Supplementary Materials. The defined disparity reduction and remaining can be estimated via methods reviewed in the next section.

## 4. Estimation Methods

In this section, we review three estimation methods for disparity reduction and remaining and provide two extensions of existing methods that can be used in practical settings.

### 4.1. Existing Methods

**Regression-Based Method.** The regression-based method was first introduced by Jackson and VanderWeele (2018). The method can be applied to a case of a single continuous mediator. Therefore, we use intervention 1 (childhood abuse) for illustrating this method. For simplicity, we only focus on the disparity between Black women ($R = 1$) and White men ($R = 0$). The specific estimation procedure for disparity reduction/remaining is as follows.

1. Fit outcome models regressed on race-gender status, baseline covariates, and additionally intermediate confounding (child SES $S$), and finally the mediator (childhood abuse $M_1$) as

$$\begin{aligned} Y &= \phi_0 + \phi_1 R + \phi_2 C + e_1, \\ Y &= \gamma_0 + \gamma_1 R + \gamma_2 S + \gamma_3 C + e_2, \text{ and} \\ Y &= \theta_0 + \theta_1 R + \theta_2 S + \theta_3 M_1 + \theta_4 C + e_3, \end{aligned} \quad (1)$$

where $\phi_1$ represents the CVH disparity between Black women and White men given baseline covariates; $\gamma_1$ represents the disparity within levels of child SES given baseline covariates; $\theta_1$ represents the disparity remaining after intervening to equalize childhood abuse across groups within the level of child SES given baseline covariates. To ensure that these regression coefficients are the effect estimates given $C = c$, covariates should be centered at $C = c$ (continuous) or be a reference group (categorical) when fitting these models.

2. Disparity reduction is estimated as $\hat{\delta}^1(1) = \hat{\gamma}_1 - \hat{\theta}_1 + (1 - \hat{\theta}_2/\hat{\gamma}_2)(\hat{\theta}_1 - \hat{\gamma}_1)$; disparity remaining is estimated as $\hat{\zeta}^1(0) = \hat{\theta}_1 + (\hat{\theta}_2/\hat{\gamma}_2)(\hat{\phi}_1 - \hat{\gamma}_1)$.

Note that $\hat{\gamma}_1 - \hat{\theta}_1$ is the disparity reduction estimate after intervening on childhood abuse within levels of child SES ($S$) given baseline covariates. This estimate is meaningful if investigators are interested in an intervention to remove disparities in childhood abuse that cannot be attributable to the disparities in child SES. However, Jackson and VanderWeele (2018) argue that this estimate within levels of child SES may not be desirable because achieving disparity reductions for children who have the same child SES is sub-optimal. For example, if we only consider those who have a high level of child SES, the disparity reduction after equalizing childhood abuse between the groups is likely underestimated compared to those across all levels of child SES. Graphically, the disparity reduction estimate within levels of child SES excludes the following path:
$R \to S \to M_1 \to (M_2, M_3) \to Y$, where the parentheses imply that the path goes both through and not through $M_2$ and $M_3$.

To obtain disparity reduction across all levels of child SES, we add $(1 - \hat{\theta}_2/\hat{\gamma}_2)(\hat{\theta}_1 - \hat{\gamma}_1)$, which is the mediated effect of child SES ($\hat{\theta}_1 - \hat{\gamma}_1$) scaled by the proportion of the mediated portion via childhood abuse $(1 - \hat{\theta}_2/\hat{\gamma}_2)$. Likewise, we add $(\hat{\theta}_2/\hat{\gamma}_2)(\hat{\phi}_1 - \hat{\gamma}_1)$ to disparity remaining after intervening on childhood abuse within levels of child SES ($\hat{\theta}_1$). Due to the added term, disparity reduction and remaining estimators differ from the conventional difference-in-coefficients approach. However, hereafter, we refer to this method as the *difference-in-coefficients* estimator to differentiate it from the new type of regression-based method that we introduced later.

Standard errors can be obtained by delta methods or bootstraps. The regression-based approach is efficient in terms of standard errors and is generally straightforward to use. However, this difference-in-coefficients estimator is not straightforward for outcome models with nonlinear relationships (e.g., interactions). In disparities research, it is common to specify differential effects of mediators on the outcome by race or gender. For example, previous literature indicates that the effect of perceived discrimination on CVH is larger for White men than Black women (Bey et al., 2019). The estimator does not allow any differential effects to be incorporated without further calculations.

In addition, the difference-in-coefficients estimator with a binary mediator/outcome results in biased estimates. With a binary outcome, the estimator generally produces biased estimates due to varying scales across nested logistic/probit regressions (for detailed information, refer to MacKinnon, Cheong, & Pirlott, 2012). With a binary mediator, caution is required since this estimator approximates the sample average of mediation effects only when the exposure distribution is symmetric (Li, Schneider, & Bennett, 2007).

**Weighting-Based Method.** Jackson (2019) developed two weighting-based estimators: ratio of mediator probability weighting (RMPW) and inverse odds ratio weighting (IORW) estimation. According to that study, the two estimators perform similarly, but IORW has a higher burden in terms of modeling perspectives as IORW requires fitting two additional exposure models. Therefore, we only address RMPW in this study. The RMPW estimator can be applied for a single categorical mediator, and thus, we use a dichotomized childhood abuse variable as a mediator for illustration. The specific estimation procedure is as follows.

1. Fit a mediator model, regressing childhood abuse on baseline covariates among White men ($R = 0$). Since childhood abuse is dichotomized, we use logistic regression. Based on this fitted model, compute the predicted probability of $M_{1i}$ given $C_i$ for each subject (i.e., $P(M_{1i}|R = 0, C_i)$).

2. Fit another mediator model, regressing childhood abuse on baseline covariates and the intermediate confounder (child SES) among Black women ($R = 1$). Based on this fitted model, compute the predicted probability of $M_{1i}$ given $S_i$ and $C_i$ for each subject (i.e., $P(M_{1i}|R = 1, S_i, C_i)$).

3. Calculate the average CVH ($Y$) among Black women given $C = c$, weighted by the ratio of the two predicted probabilities as $W_i = \frac{P(M_{1i}|R=0,C_i)}{P(M_{1i}|R=1,S_i,C_i)}$. This estimates $E[WY|R = 1, C = c]$. This quantity is obtained as the intercept in a weighted regression of $Y$ on $C$ among individuals with $R = 1$. Covariates should be centered at $C = c$ or be a reference group when fitting this regression model.

4. Calculate the average CVH for Black women and White men given $C = c$ (i.e., $E[Y|R = 1, c]$ and $E[Y|R = 0, c]$). Likewise, these quantities are obtained as the intercept in a regression of $Y$ on $C$ among individuals with $R = 1$ and $R = 0$, respectively.

5. Disparity reduction is estimated as $\hat{\delta}^1(c) = \hat{E}[Y|R = 1, c] - \hat{E}[\hat{W}Y|R = 1, c]$ and disparity remaining is estimated as $\hat{\zeta}^1(c) = \hat{E}[\hat{W}Y|R = 1, c] - \hat{E}[Y|R = 0, c]$.

Refer to Jackson (2019) for marginal effects that are averaged over covariates. Standard errors can be obtained from bootstraps.

One advantage of this RMPW estimator is its flexibility to accommodate nonlinear terms since the estimator does not change regardless of the fitted models. In modeling perspectives, the estimator is also advantageous since it requires fitting two different mediator models only. Disadvantages of this estimator, however, include that it only addresses categorical mediators since most weighting-based approaches do not work very well with continuous variables. Also, weighting-based approaches are generally less efficient in terms of standard errors compared to regression-based approaches (VanderWeele, 2010a).

**Imputation-Based Method.** The imputation-based method is developed by Park et al. (2020) to address the case of intervening on multiple mediators that are continuous or categorical. Therefore, we use intervention 2 (education and discrimination) to illustrate this method. The specific estimation procedure is as follows.

1. Fit confounder models, regressing each confounder (childhood SES and child abuse) on the race-gender group and baseline covariates as $\psi_r(c) \equiv p(S_i, M_{1i}|R_i = r, C_i = c)$.

Based on the fitted models, we compute a predicted value of confounders for each subject (denoted as $\tilde{s}_i$ and $\tilde{m}_{1i}$) among White men ($R = 0$), after forcing $R = 1$.

2. Fit an outcome model, regressing CVH on the race-gender group, intermediate confounders, mediators, and baseline covariates as $\mu_{rsm_1m_2m_3}(c) \equiv E(Y_i | R = r, S_i = s, M_{1i} = m_1, M_{2i} = m_2, M_{3i} = m_3, C_i = c)$. Based on the fitted outcome model, compute a predicted outcome value for each subject among White men ($R = 0$), after forcing $R = 1$, and imputing $\tilde{s}_i$ and $\tilde{m}_{1i}$ as $\mu_{1\tilde{s}_i\tilde{m}_{1i}M_{2i}M_{3i}}(C_i)$.

3. The predicted outcome values obtained from step 2 will be average over $i$ among White men given $C = c$. This computes 
$E[Y(G_{M_{2,3}|c}(0))|R = 1, c] = \frac{1}{n_0} \sum_{i \in \Pi_0} \mu_{1\tilde{s}_i\tilde{m}_{1i}M_{2i}M_{3i}}(c)$, where $\Pi_0$ is the subjects (of size $n_0$) in group $R = 0$.

4. The disparity reduction is estimated as 
$\hat{\delta}^2(0) = \hat{E}[Y|R = 1, c] - \hat{E}[Y(G_{M_{2,3}|c}(0))|R = 1, c]$ and disparity remaining is estimated as $\hat{\zeta}^2(0) = \hat{E}[Y(G_{M_{2,3}|c}(0))|R = 1, c] - \hat{E}[Y|R = 0, c]$.

Refer to Park et al. (2020) for marginal effects that are averaged over covariates. Standard errors can be obtained from bootstraps.

To differentiate this estimator with the new imputation method introduced later, we refer to this method as the *multiple-mediator-imputation* estimator. This multiple-mediator-imputation estimator is perhaps the most flexible among the existing estimators since it can address 1) any nonlinear terms, 2) multiple mediators and a single mediator, and 3) different variable types of confounders and mediators. However, depending on the causal structure of variables, there could be more burden in correctly specifying models than regression- or weighting-based methods. This estimator requires fitting models for intermediate confounders instead of mediators. Therefore, this estimator is advantageous (in terms of modeling burden) only when the number of mediators exceeds or equals the number of intermediate confounders.

**4.2. Two Extensions**

**Product-of-Coefficients Method.** While the difference-in-coefficients estimator is straightforward to use, it is less useful when differential effects of the mediator exist by groups. Differential effects by groups are common in disparities research, so assuming constant mediator effects across groups may be too restrictive in many empirical settings. Furthermore, the difference-in-coefficients estimator cannot address a categorical mediator or outcome.

Therefore, a new type of regression-based approach is introduced here using a product of two coefficients: one from a mediator model and another from an outcome model. This model will be referred as the *product-of-coefficients* estimator although the estimator differs from the one often used in conventional mediation analysis. This product-of-coefficients estimator allows a differential effect of mediators by groups and can be easily modified to address a categorical mediator. Still, the simplicity of a regression-based approach is maintained. A specific estimation procedure using intervention 1 is as follows.

1. After centering $C = c$, fit a mediator and outcome model as

$$\begin{aligned} M &= \alpha_0 + \alpha_1 R + \alpha_2 C + e_m, \\ Y &= \beta_0 + \beta_1 R + \beta_2 S + \beta_3 M_1 + \beta_4 R M_1 + \beta_5 C + e_y. \end{aligned} \quad (2)$$

2. The disparity reduction is estimated as $\hat{\delta}^1(c) = \hat{\alpha}_1 \times (\hat{\beta}_3 + \hat{\beta}_4)$ and the disparity remaining is estimated as $\hat{\zeta}^1(c) = \hat{\phi}_1 - \hat{\alpha}_1 \times (\hat{\beta}_3 + \hat{\beta}_4)$, where $\hat{\phi}_1$ is the initial disparity given covariates. Alternatively, if investigators are willing to additionally model intermediate confounders, $\hat{\zeta}^1(c) = \hat{\beta}_1 + \hat{\beta}_2 \hat{\kappa}_1 + \hat{\beta}_4 \hat{\alpha}_0$, where $\hat{\kappa}_1$ is the average disparity in child SES ($S$) between Black women and White men ($R = 1$ and $R = 0$, respectively) given the covariates. A proof for this estimator is shown in Appendix A.

3. For a categorical mediator, $\alpha_0$ is no longer the average mean of $M$, but the average probability of $M = m$ for White men given covariates; $\alpha_1$ is no longer the average disparity in $M$, but the average probability difference in $M = m$ between Black women and White men given covariates.

Note that disparity reduction is the product of $\hat{\alpha}_1$ and $\hat{\beta}_3 + \hat{\beta}_4$. Here, $\hat{\alpha}_1$ is the average disparity in childhood abuse between Black women and White men given baseline covariates. In conventional mediation analysis, instead of $\hat{\alpha}_1$, the average disparity in childhood abuse was used within levels of child SES given baseline covariates. This conventional estimator is interested in removing disparities in childhood abuse that cannot be attributable to the disparities in child SES. In contrast, we are interested in removing disparities in childhood abuse across all levels of child SES.

Here, we present two ways of estimating remaining disparity. In modeling perspectives, the alternative estimator is less advantageous than the original estimator since we must additionally model intermediate confounders. Later, our simulation study shows that the original estimator only works for a continuous mediator while the alternative estimator works for continuous or categorical mediators.

One disadvantage of this product-of-coefficients estimator is that it cannot address a categorical outcome and nonlinear terms other than exposure-mediator interactions, for which either a weighting or imputation method should be considered. Still, this estimator has advantages over the difference-in-coefficients estimator because a categorical mediator and exposure-mediator interactions can be addressed.

**Single-Mediator-Imputation Method.** While the multiple-mediator-imputation estimator is flexible enough to accommodate both multiple mediators and a single mediator, the estimator may not be necessarily advantageous for a single mediator's case in modeling perspectives. The estimator does not require researchers to model mediators but, instead, intermediate confounders. Therefore, the estimator is beneficial when the number of mediators exceeds the number of intermediate confounders. In the case of a single mediator, the estimator is no longer beneficial if there is more than one intermediate confounder (because the number of mediator is one), which is likely in many settings.

To alleviate burdens of modeling many intermediate confounders, we present a modified imputation-based approach designed for a single mediator. This new *single-mediator-imputation* estimator can address the single mediator's case without the burden of specifying extra confounder models while maintaining its benefit of flexibility in addressing nonlinear terms and different variable types of mediators and outcome. We illustrate this modified approach with intervention 1 (childhood abuse).

1. Fit a mediator model, regressing the mediator (childhood abuse) on the race-gender group and baseline covariates as $\phi_r(c) \equiv p(M_{1i}|R_i = r, C_i = c)$. Based on the fitted model, we compute the predicted value of the mediator for each subject (denoted as $\tilde{m}_{1i}$) among Black women ($R = 1$), after forcing $R = 0$.

2. Fit an outcome model, regressing CVH on the race-gender group, intermediate confounder, mediator, and baseline covariates as
   $\mu_{rsm}(c) \equiv E(Y_i|R_i = r, S_i = s, M_{1i} = m_{1i}, C_i = c)$. Based on the fitted model, compute a predicted outcome value for each subject among Black women after imputing $\tilde{m}_{1i}$ as $\mu_{1S_i\tilde{m}_{1i}}(c)$.

3. The predicted outcome values obtained from step 2 will be averaged over $i$ among Black women given $C = c$. This computes
   $E[Y(G_{M_1|c}(0))|R = 1, c] = \frac{1}{n_1} \sum_{i \in \Pi_1} \mu_{1S_i\tilde{m}_{1i}}(c)$, where $\Pi_1$ is the subjects (of size $n_1$) in group $R = 1$.

4. The disparity reduction is estimated as
   $\hat{\delta}^1(c) = \hat{E}[Y|R = 1, c] - \hat{E}[Y(G_{M_1|c}(0))|R = 1, c]$ and disparity remaining is estimated as $\hat{\zeta}^1(c) = \hat{E}[Y(G_{M_1|c}(0))|R = 1, c] - \hat{E}[Y|R = 0, c]$. A proof for this estimator is shown in Appendix B.

We only present disparity reduction/remaining conditional on covariates here. However, we provide a proof for marginal effects in Appendix C. Note that models for a mediator and an outcome were fitted as in the product-of-coefficients estimator for a continuous mediator. While the same burden of modeling is required between the two estimators, this single-mediator-imputation estimator can address any nonlinear terms while the product-of-coefficients estimator can address only the exposure-mediator interaction. Standard errors can be obtained via bootstraps.

## 5. Simulation Study

This section compares the performances of the estimation methods under different conditions through a small but thorough simulation study. This simulation study focuses on intervening on a single mediator in which multiple estimation methods are available. For simplicity, we refer to difference-in-coefficients, product-of-coefficients, RMPW, single-mediator-imputation, and multiple-mediator-imputation as estimators 1, 2, 3, 4, and 5, respectively, in this section.

**Data Generation.** The procedure for data generation is as follows. First, a binary treatment $R$ is created, which takes the value of 0 or 1, with the probability of 0.5 for randomly assigning the value for each observation. For observations with $R = 1$, a covariate $C$ is generated from a truncated normal distribution with mean 50 and standard deviation 12 within the interval $(25, 75)$; For observations with $R = 0$, a covariate $C$ is generated from the same distribution but with the shifted mean to 48. The generated $C$ is dichotomized by 50, taking the value of 1 or 2. We then create $S$, $M$, and $Y$ using equations (3), (4), and (5), respectively, with added error terms from a standard normal distribution. To generate synthetic data that mimics real data, we use the relationship between variables in the MIDUS data from Section 2 to create $S$, $M$, and $Y$.

$$S = a_0 + a_1 R + a_2 C + e_s \tag{3}$$

$$M = b_0 + b_1 R + b_2 C + b_3 S + e_m \tag{4}$$

$$Y = c_0 + c_1 R + c_2 S + c_3 M + c_4 RM + c_5 C + e_y \tag{5}$$

For a binary mediator case, the same procedure is used, but we use an $M$ dichotomized by its median to fit the logistic regression of (4). Then a binary mediator is generated for the synthetic data that takes the value of 0 or 1 with the probability $\frac{\exp(b_0+b_1R+b_2C+b_3S)}{1+\exp(b_0+b_1R+b_2C+b_3S)}$ for being $M = 1$. The true values for each scenario are shown in Table 2. We fixed the percentage of disparity reduction as 30% across different settings to ensure comparability.

**Simulation Setting.** We use a binary and continuous mediator since the performance of the estimation methods may depend on the variable type. For each type of mediator, we consider three sample sizes $n = \{100, 500, 1000\}$, which cover reasonably small and large sample sizes. For each fixed sample size, an important condition that we vary is the ratio between the $R - M$ and $M - Y$ association, which is computed as

$$r = \frac{|E(M|R=0, C=c) - E(M|R=1, C=c)|}{|E(Y|R=1, S=s, M=1, C=c) - E(Y|R=1, S=s, M=0, C=c)|},$$

If some estimation methods are sensitive to this ratio, its performance would change as the ratio varies. We here consider $r = \{0.3, 0.5, 1, 2, 3\}$, which covers low and high ratios. As this ratio increases, the $R - M$ association increases compared to the $M - Y$ association. Thus, we consider 15 scenarios with different $n$ and $r$ values for each type of mediator. To set the desired level of the ratio $r$, we change the coefficients. Table 2 shows how to set the coefficients for each scenario. Other than the three coefficient values in the table, the remaining coefficient values are fixed as the coefficients from the MIDUS data.

**Performance Metrics.** In this study, the following metrics are used for performance comparison between the estimation methods: bias, the root mean square errors (RMSE), and 95% confidence interval coverage using the percentile bootstrap method (Efron, 1982) with the number of bootstrap replicates $B = 1000$. For each scenario, we make $M = 1000$ replicates of sample, and the performances are averaged over the 1000 repetitions. Let $\hat{\delta}_{bm}(c)$ and $\hat{\zeta}_{bm}(c)$ denote the estimate of the disparity reduction and disparity remaining from the $b$th bootstrap sample of the $m$th sample replicate.

Table 2

*Coefficient values for each scenario and corresponding parameters*

| Mediator type | Ratio | Coefficients | | | True effects | |
|---|---|---|---|---|---|---|
| | | $b_1$ | $c_1$ | $c_3$ | $\delta^1(1)$ | $\zeta^1(0)$ |
| Continuous | 0.3 | 0.244 | -1.048 | -1.322 | -0.263 | -0.610 |
| | 0.5 | 0.326 | -1.048 | -1.112 | -0.262 | -0.611 |
| | 1 | 0.477 | -1.048 | -0.900 | -0.263 | -0.611 |
| | 2 | 0.690 | -1.049 | -0.750 | -0.263 | -0.612 |
| | 3 | 0.852 | -1.049 | -0.684 | -0.263 | -0.612 |
| Binary | 0.3 | 1.189 | -0.644 | -1.327 | -0.283 | -0.660 |
| | 0.5 | 1.446 | -0.674 | -1.126 | -0.297 | -0.689 |
| | 1 | 1.993 | -0.709 | -0.913 | -0.311 | -0.724 |
| | 2 | 0.388 | -0.109 | -0.518 | -0.053 | -0.124 |
| | 3 | 0.333 | -0.087 | -0.475 | -0.044 | -0.102 |

Note. 1) $b_1, c_1$, and $c_4$: regression coefficients from (4) and (5).

1. The biases:

$$\frac{1}{M}\sum_{m=1}^{M}\left(\delta(c) - \hat{\delta}_{1m}(c)\right) \quad \text{and} \quad \frac{1}{M}\sum_{m=1}^{M}\left(\zeta(c) - \hat{\zeta}_{1m}(c)\right)$$

2. The root mean square errors (RMSEs):

$$\sqrt{\frac{1}{M}\sum_{m=1}^{M}\left(\delta(c) - \hat{\delta}_{1m}(c)\right)^2} \quad \text{and} \quad \sqrt{\frac{1}{M}\sum_{m=1}^{M}\left(\zeta(c) - \hat{\zeta}_{1m}(c)\right)^2},$$

where $\delta(c)$ and $\zeta(c)$ are the true disparity reduction and remaining from population data, respectively.

3. The 95% confidence interval coverage:

$$\frac{1}{M}\sum_{m=1}^{M}I\left(\hat{\delta}_m^L < \delta(c) < \hat{\delta}_m^U\right) \quad \text{and} \quad \frac{1}{M}\sum_{m=1}^{M}I\left(\hat{\zeta}_m^L < \delta(c) < \hat{\zeta}_m^U\right)$$

where $(\hat{\delta}_m^L, \hat{\delta}_m^U)$ and $(\hat{\zeta}_m^L, \hat{\zeta}_m^U)$ are the 95% bootstrap confidence interval for the disparity reduction and disparity remaining of the $m$th sample replicate, respectively.

**Simulation Results.** The simulation results for a continuous and binary mediator are summarized in Figures 2 and 3, respectively. The exact numerical values of the performance metrics can also be found in Supplementary Materials. From Figure 2, the first thing to notice is estimator 1's poor performance compared to the others. It provides a biased estimate, and the coverage rate does not reach the nominal level even with a sample size of 1000. This poor performance is not surprising given that estimator 1 cannot address any nonlinear terms.

Except for estimator 1, all estimators perform well with a sample size of 1000 and a ratio of 2 or higher. However, with smaller sample sizes or a ratio less than 2, the estimators do perform differently. With a sample size of 100 and 500, estimators 2 and 5 perform equally well regarding bias and coverage, regardless of ratios. Yet, estimator 2 demonstrates a smaller RMSE. For example, RMSEs for disparity reduction obtained from estimators 2 and 5 (with a sample size of 100 and a ratio of 0.3) are 0.202 and 0.308, respectively. This implies that estimator 2 performs the best for a continuous mediator in terms of bias and variance.

Unlike estimators 2 and 5, estimator 4 shows a high coverage rate when the ratio is less than 2. Even with a sample size of 1000, the coverage rate for disparity reduction exceeds 0.98 with a ratio less than 2. This implies that estimator 4 is inefficient in terms of standard errors (here, shown as wide confidence intervals) when the exposure-mediator association is less than twice the mediator-outcome association.

Again, from Figure 3, we observe a biased result for estimator 1. As discussed in Section 4, estimator 1 cannot address a binary mediator and any nonlinear relationships. Except for estimator 1, all estimators for disparity reduction perform well with a sample size of 1000 and a ratio of 1 or higher. With smaller sample sizes ($n = 100$ and 500), estimators 2, 4, and 5 perform equally well in terms of RMSE and coverage. Yet, estimator 2 demonstrates the smaller bias for disparity reduction than the other estimators.

One noteworthy thing is a biased result of estimator 2 for disparity remaining. The

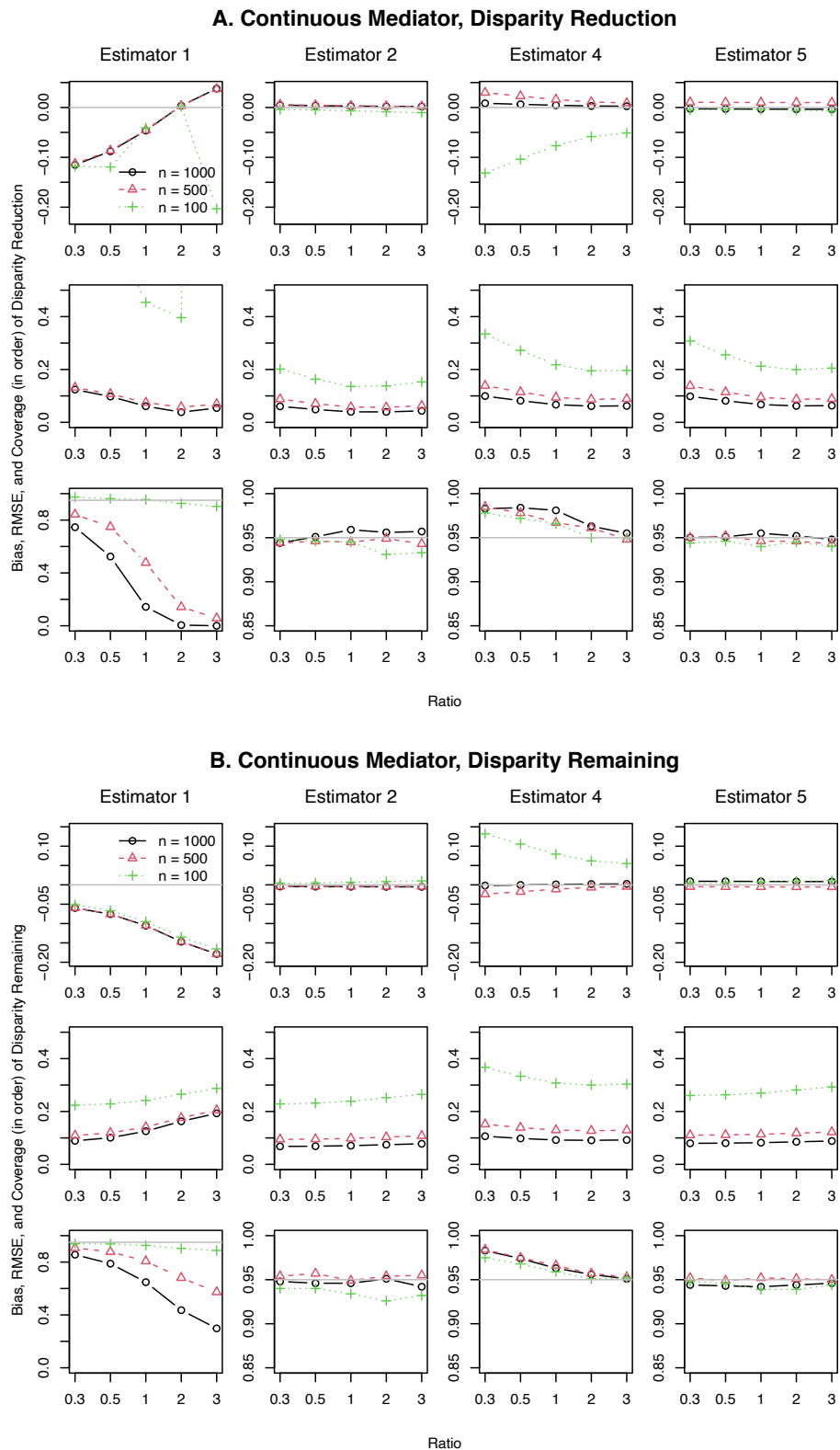

*Figure 2.* Performances of disparity reduction (A) and disparity remaining (B) with a continuous mediator.

Note. 1)1: Difference-in-coefficients estimator; 2: Product-of-coefficients estimator; 4: Single-mediator-imputation estimator; 5: multiple-mediator-imputation. 2) Estimator 3 is not considered since it is only available for a binary mediator.

ESTIMATORS FOR CAUSAL DECOMPOSITION 23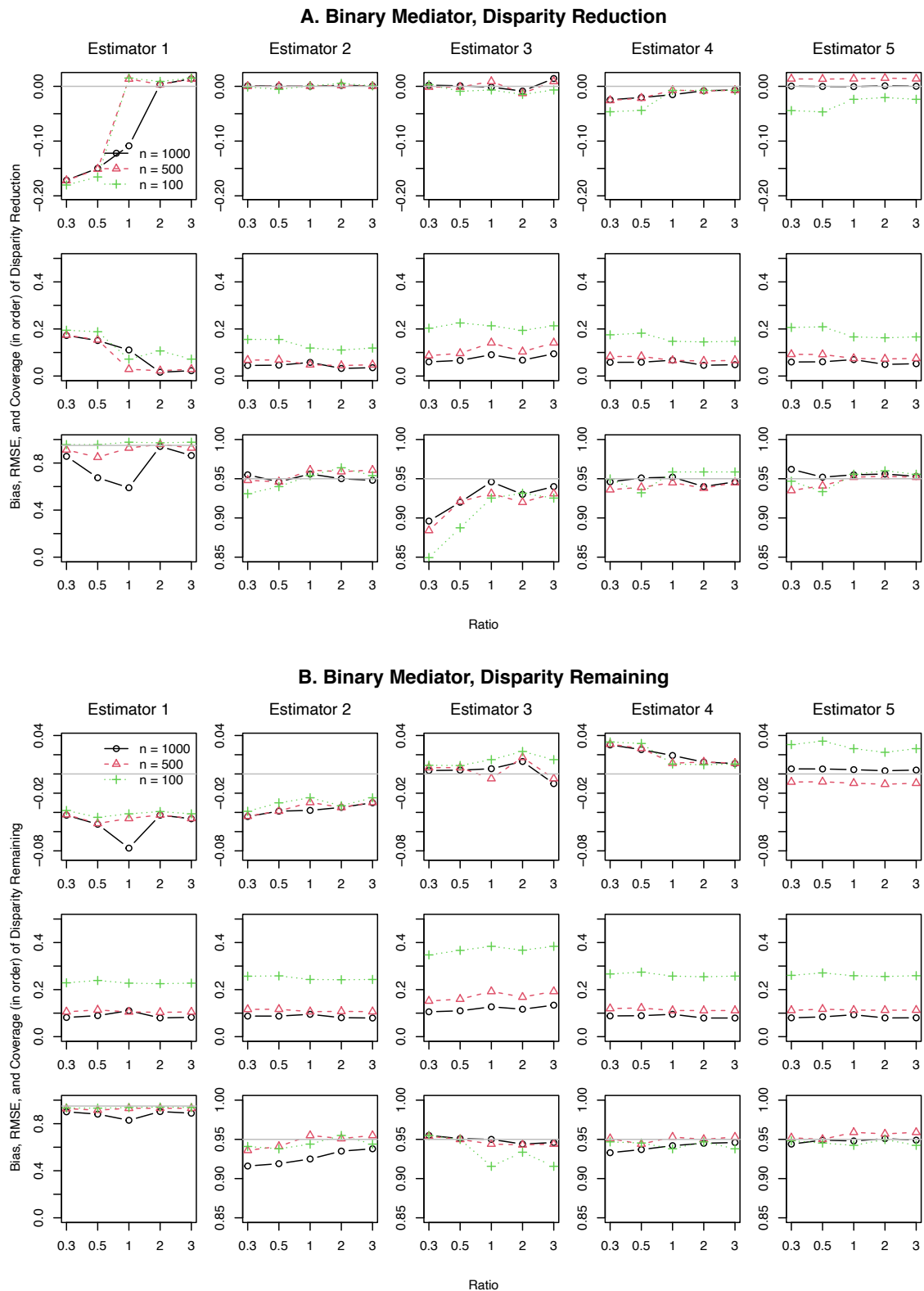

*Figure 3.* Performances of disparity reduction (A) and disparity remaining (B) with a binary mediator.

Note. 1) 1: Difference-in-coefficients estimator; 2: Product-of-coefficients estimator; 3: RMPW estimator; 4: Single-mediator-imputation estimator; 5: multiple-mediator-imputation.

disparity remaining estimate is biased with the original estimator $(\hat{\zeta}(0) = \hat{\tau}(1,0) - \hat{\delta}(1))$. In contrast, the estimate is unbiased with the alternative estimator, as shown in Figure 4. By additionally modeling intermediate confounders, estimator 2 demonstrates superior performance compared to the other estimators. This result suggests that the alternative estimator for disparity remaining should be used for a binary mediator.

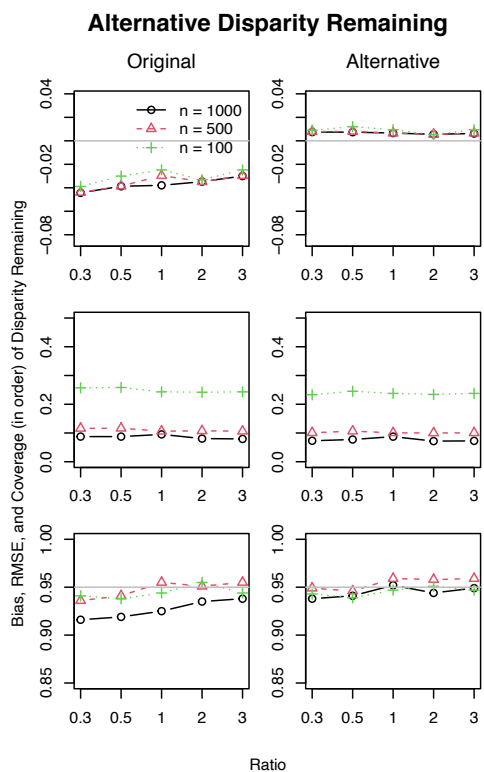

*Figure 4*. Performances of estimator 2 between original and alternative disparity remaining estimate with a binary mediator.

Note. 1) The left three panels are the same as the panels under Estimator 2 in Figure 3 and shown here for reference.

Another thing to note is a low coverage rate for estimator 3 when the ratio is less than 1. When the exposure-mediator association is less than the mediator-outcome association, the coverage rate for estimator 3 decreases dramatically. For example, with a sample size of 100 and a ratio of 0.3, the coverage rate is 0.85. This low coverage rate is due to narrower confidence intervals than the nominal level, potentially leading to a wrong conclusion.

# 6. Application

**Choosing Between Methods**

Based on our review of the methods and the simulation study, we provide recommendations for selecting an optimal method. We illustrate the practice of choosing an optimal method through reanalysis of Lee et al. (2021). The availability of estimation methods depends on the variable types of the mediator and outcome, the number of mediators, and the modeling assumptions that researchers are willing to make (e.g., nonlinearity). Therefore, the first step is to check which methods are available given the research questions and data at hand. Table 3 shows the summary of available estimation methods depending on conditions. From the table, we note that the difference-in-coefficients method is the most restrictive while the multiple-mediator-imputation method is the most flexible.

For intervention 1, our research question is to what extent the CVH disparity would be reduced if we reduce the exposure to childhood abuse for Black women to the level of White men. The mediator is childhood abuse (mean = 1.44, SD = 0.62) and the outcome is CVH, where higher values indicate better CVH (mean = 8.09, SD = 2.12). We model differential effects of childhood abuse between Black women and White men (interaction between $R$ and $M_1 = 0.52$, $p < 0.03$). Given these conditions, the following methods are available: product-of-coefficients, single-mediator-imputation, and multiple-mediator-imputation (estimators 2, 4, and 5 from Table 3). Normally, the multiple-mediator-imputation method may not be the best choice for a single mediator's case due to the burden of modeling intermediate confounders. However, in the context of our example, there is only one intermediate confounder (child SES) for intervention 1. Therefore, the multiple-mediator-imputation method is still a good option.

One question that remains is which method should be used among these multiple options. Suppose a software package supports the application of all methods (although this is not true). In that case, one should consider both the bias and efficiency of estimators

Table 3

*Summary of available methods*

| Approach | Estimator | Type of Mediator | | Type of Outcome | | Number of Mediators | | Nonlinear terms | | |
|---|---|---|---|---|---|---|---|---|---|---|
| | | Cat. | Cont. | Cat. | Cont. | Single | Multiple | No | R×M | Others |
| Regression | 1 | | ✓ | | ✓ | ✓ | | ✓ | | |
| | 2 | ✓ | ✓ | | ✓ | ✓ | | ✓ | ✓ | |
| Weighting | 3 | ✓ | | ✓ | ✓ | ✓ | | ✓ | ✓ | ✓ |
| Imputation | 4 | ✓ | ✓ | ✓ | ✓ | ✓ | | ✓ | ✓ | ✓ |
| | 5 | ✓ | ✓ | ✓ | ✓ | ✓ | ✓ | ✓ | ✓ | ✓ |

Note. 1) 1: Difference-in-coefficients, 2: Product-of-coefficients, 3: RMPW, 4: Single-mediator-imputation, and 5: multiple-mediator-imputation. 2) Cat.=Categorical, Cont.=Continuous, and R×M: Differential effects of mediators by groups.

across the methods given the sample size and ratio between the exposure-mediator and mediator-outcome association. In our case, the sample size is 1978, and the ratio is 1.76. The simulation study suggests that the product-of-coefficients method performs the best for a continuous mediator in terms of bias and variance. However, this superior performance of the product-of-coefficients method is due to its modeling assumption that no other nonlinear terms are allowed except for the exposure-mediator interaction. If other nonlinear terms are modeled in the mediator or outcome model, one should consider using one of the imputation methods. Given the sample size and the ratio, both the single-mediator-imputation and the multiple-mediator-imputation methods are expected to work well.

Although the mediator (childhood abuse) is continuous, we consider a binary mediator's case for illustration by dichotomizing mediator by its median. Given the same condition as before, the following methods are available: product-of-coefficients, RMPW, single-mediator-imputation, and multiple-mediator-imputation (2, 3, 4, and 5 from Table 3). The ratio is 0.17 after dichotomizing the mediator. Given this ratio, the simulation study suggests that the product-of-coefficients (using the alternative estimator for disparity remaining) and the imputation methods should work well. If investigators are willing to assume no other nonlinear terms except for the exposure-mediator interaction, the product-of-coefficients method should be considered. If other nonlinear terms are modeled, the imputation methods should be considered. While the RMPW method is also an available option, caution is required as the confidence interval obtained from nonparametric bootstraps may be narrower than expected for a ratio less than 1.

For intervention 2, our research question is to what extent the CVH disparity would be reduced if we decrease discrimination and increase education for Black women to the level of White men. The mediators are discrimination (standardized mean $= -0.01$, SD $= 0.99$) and education (mean $= 7.72$, SD $= 2.53$), and the outcome is CVH. We model differential effects of discrimination (interaction between $R$ and $M_2 = 0.10$, $p = 0.47$). Although the interaction effect is not significant in our sample, the differential effect of discrimination by race and gender is known from previous literature (Bey et al., 2019). The only available method for multiple mediators is the multiple-mediator-imputation method.

**Summary of Findings**

The estimates for disparity reduction and remaining obtained from different estimation methods are shown in Table 4. We begin by noting that the initial disparity for Black women compared to White men is $\tau(1, 0) = -0.97$, with the confidence interval bounded away from zero, which means that Black women's CVH is worse (unhealthier) than White men among those who have the average level of age and genetic vulnerability. The initial disparity is slightly smaller for the regression-based method ($\tau(1, 0) = -0.93$). Once the disparity is observed, social scientists would also want to know how to reduce the disparity, for example, by reducing Black women's exposure to childhood abuse to the level of White men.

The first bracket of Table 4 shows disparity reduction and remaining due to intervening on childhood abuse. The estimand $\delta^1(1)$ ranges between $-0.07$ (for using Estimator 2) and $-0.12$ (for using Estimator 4 or 5) and the confidence intervals (for using all three estimators) cover zero. Given the assumptions, this means that the CVH disparity between Black women and White men won't be significantly reduced even if we intervene to decrease Black women's childhood abuse to the level of White men.

The second bracket of Table 4 shows disparity reduction and remaining when the mediator (childhood abuse) was binary. The estimand $\delta^1(1)$ ranges between 0.01 (Estimator 2) and $-0.09$ (Estimator 4), and the confidence intervals cover zero. The interpretation for $\delta^1(1)$ is the same as the continuous mediator case. One thing to note is that compared to the confidence interval (CI $= -0.147, -0.026$) of the disparity reduction for the product-of-coefficients method, the CI from the weighting method $(-0.042, -0.048)$ is narrower while the CI from the single-mediator-imputation method $(-0.216, -0.051)$ is wider. This result is consistent with the simulation result when the ratio is less than 1.

The third bracket of Table 4 shows disparity reduction and remaining due to intervening on education and perceived discrimination simultaneously. According to the multiple-mediator-imputation method, disparity reduction is $\delta^2(1) = 0.52$, with the confidence interval bounded away from zero. This implies that CVH disparity between Black women and White men will be significantly reduced if we intervene to decrease discrimination and increase education for Black women to the level of White men. The percentage reduction due to this intervention is 54.3%.

In this example, the same conclusion is derived from different estimation methods. Yet, it is important to note that a different conclusion could be derived depending on estimation methods, particularly when a sample size is small or the exposure-mediator association is smaller than the mediator-outcome association.

## 7. Discussion

Estimation of disparity reduction/remaining is an important topic in causal decomposition analysis. Unlike causal mediation analysis based on natural indirect effects[2]

---

[2] The outcome difference in response to a change in a mediator naturally occurs under one condition versus another.

Table 4

*Estimates of the disparity reduction and disparity remaining for Black women vs. White men*

|  | Estimate (95% CI) | | |
|---|---|---|---|
| Intervention 1 (continuous) | Estimator 2 | Estimator 4 | Estimator 5 |
|     Initial disparity ($\tau^1(1,0)$) | -0.927 | -0.965 | -0.965 |
|     (95% CI) | (-1.236, -0.636) | (-1.262, -0.647) | (-1.262, -0.647) |
|     Disparity remaining ($\zeta^1(0)$) | -0.855 | -0.847 | -0.847 |
|     (95% CI) | (-1.153, -0.564) | (-1.149, -0.536) | (-1.147 -0.535) |
|     Disparity reduction ($\delta^1(1)$) | -0.071 | -0.118 | -0.118 |
|     (95% CI) | (-0.182, 0.030 ) | (-0.289, 0.042) | (-0.291 0.046) |
|     % reduction | 7.7% | 12.2% | 12.2% |
| Intervention 1 (binary) | Estimator 2 | Estimator 3 | Estimator 4 |
|     Initial disparity ($\tau^1(1,0)$) | -0.874 | -0.965 | -0.965 |
|     (95% CI) | (-1.191, -0.558) | (-1.294, -0.672) | (-1.258, -0.659) |
|     Disparity remaining ($\zeta^1(0)$) | -0.883 | -0.932 | -0.878 |
|     (95% CI) | (-1.192, -0.559) | (-1.288, -0.673 ) | (-1.151, -0.585) |
|     Disparity reduction ($\delta^1(1)$) | 0.008 | -0.033 | -0.087 |
|     (95% CI) | (-0.121, 0.120) | (-0.042, 0.048) | (-0.216, 0.051) |
|     % reduction | -0.9% | 3.4% | 9.0% |
| Intervention 2 | Estimator 5 | | |
|     Initial disparity ($\tau^2(1,0)$) | -0.965 | | |
|     (95% CI) | (-1.255, -0.658 ) | | |
|     Disparity remaining ($\zeta^2(0)$) | -0.441 | | |
|     (95% CI) | ( -0.812, -0.063) | | |
|     Disparity reduction ($\delta^2(1)$) | -0.524 | | |
|     (95% CI) | ( -0.821 , -0.223) | | |
|     % reduction | 54.3% | | |

Note. 1) 1: Difference-in-coefficients estimator; 2: Product-of-coefficients estimator; 3: RMPW estimator; 4: Single-mediator-imputation estimator; 5: multiple-mediator-imputation estimator. 2) R×M: Differential effects of a mediator by groups. 3) CI = confidence interval. 4) Baseline covariates are mean-centered.

(Pearl, 2009; VanderWeele, 2010b), causal decomposition analysis allows intermediate confounding, which is a substantial advantage considering that no intermediate confounding is a strong assumption that is hard to be met in many empirical settings. However, allowing intermediate confounders adds a modeling burden, since the identification result for disparity reduction/remaining depends on the distribution of intermediate confounders. Therefore, it is important to develop an estimation method that reduces the modeling burden while maintaining good performance in terms of bias and efficiency. Our newly developed product-of-coefficients estimator is flexible enough to address a categorical mediator and interaction term between the exposure and mediator. Another single-mediator-imputation estimator is flexible to address a categorical mediator/outcome and any nonlinear terms. Both estimators require modeling a mediator and an outcome (additionally modeling confounders for the product-of-coefficient estimator with a binary mediator). The product-of-coefficients estimator performs the best assuming no other nonlinear terms, except for the exposure-mediator interaction.

It is of substantive interest in choosing an optimal estimation method when the results differ when using different methods. Examining the performance of estimation methods through simulation studies provides helpful information. With a large sample size ($n \geq 1000$) and a high ratio of the exposure-mediator association to the mediator-outcome association ($r \geq 2$), all estimators perform well in terms of bias, RMSE, and coverage rate with either a binary or continuous mediator. Their performance does differ, however, with smaller sample sizes or a ratio that is less than 2. A low coverage rate of the weighting method obtained from nonparametric bootstraps with ratios less than 1 is particularly worrisome since it could inflate the type I error rate. An alternative way to calculate correct standard errors, such as that shown in Bein et al. (2018), may be necessary. It is also noteworthy that the single-mediator-imputation method provides a high coverage rate for a continuous mediator when the ratio is less than 2. Given research questions and data, investigators can use this information in choosing an optimal method for their study.

One important condition we vary in our simulation study is the ratio between the exposure-mediator and mediator-outcome association. An estimator relying heavily on modeling a mediator (such as RMPW or single-mediation-imputation) performs poorly when the exposure-mediator association is weaker than the mediator-outcome association. In contrast, the multiple-mediator-imputation relies on modeling confounders, and thus the performance does not depend on the ratio. The product-of-coefficient estimator relies on modeling a mediator but with a strong modeling assumption (i.e., no other nonlinear term other than the exposure-mediator interaction). Given this assumption, the performance of the product-of-coefficient estimator does not depend on the ratio. The ratio has been neglected in many mediation simulation studies. Yet, it appears that the ratio is an important condition to consider when assessing the performances of mediation estimators in future studies.

It is important to note the limitations of the current study. This study only addresses one way of defining disparity reduction/remaining. A different definition of disparity reduction/remaining exist (Jackson, 2019; Jackson & VanderWeele, 2018) and the performance of estimation methods for different definitions is unknown. Therefore, the simulation study can be extended to an alternative definition of disparity reduction/remaining. In addition, the current study only addresses the issues of estimating disparity reduction/remaining when the identification assumptions are met. However, the assumptions are strong, and thus, may not be met in many empirical settings. Therefore, an important future study includes developing a sensitivity analysis to possible violations of the assumptions.

**Appendix A. A Proof for the Product-of-Coefficients Estimator**

Disparity reduction for intervention 1 is defined as

$$\delta^1(1) = E[Y|R=1,c] - \sum_{s,m_1} E[Y|R=1,s,m_1,c]P(s|R=1,c)P(m_1|R=0,c), \quad (6)$$

where $s \in \mathcal{S}, m_1 \in \mathcal{M}_1$, and $c \in \mathcal{C}$. Here, $E[Y|R=1,c]$ can be rewritten as

$$\begin{aligned}
E[Y|R=1,c] &= \sum_s E[Y|R=1,s,c]P(s|R=r,c) \\
&= \sum_{s,m_1} E[Y|R=1,s,m_1,c]P(s|R=1,c)P(m_1|R=1,s,c) \\
&= \sum_{s,m_1} (\beta_0 + \beta_1 + \beta_2 s + \beta_3 m_1 + \beta_4 m_1 + \beta_5 c)P(s|R=1,c)P(m_1|R=1,s,c) \\
&= \beta_0 + \beta_1 + \beta_2 E[S|R=1,c] + \beta_3 E[M_1|R=1,c] + \beta_4 E[M_1|R=1,c] + \beta_5 c
\end{aligned} \quad (7)$$

The first equality is due to the law of total probability with respect to $S=s$. The second equality is due to the law of total probability with respect to $M_1 = m_1$. The third equality is after incorporating the outcome model. Likewise,

$$\begin{aligned}
&\sum_{s,m_1} E[Y|R=1,s,m_1,c]P(s|R=1,c)P(m_1|R=0,c) \\
&= \sum_{s,m_1} (\beta_0 + \beta_1 + \beta_2 s + \beta_3 m_1 + \beta_4 m_1 + \beta_5 c)P(s|R=1,c)P(m_1|R=0,c) \\
&= \beta_0 + \beta_1 + \beta_2 E[S|R=1,c] + \beta_3 E[M_1|R=0,c] + \beta_4 E[M_1|R=0,c] + \beta_5 c
\end{aligned} \quad (8)$$

Given equations (7) and (8), the disparity reduction can be rewritten as

$$\begin{aligned}
\delta^1(1) &= (E[M_1|R=1,c] - E[M_1|R=0,c]) \times (\beta_3 + \beta_4) \\
&= \alpha_1 \times (\beta_3 + \beta_4)
\end{aligned} \quad (9)$$

The second equality is after incorporating the mediator model.

The disparity remaining is estimated as $\zeta^1(c) = \phi - \alpha_1 \times (\beta_3 + \beta_4)$ since disparity reduction and remaining sum up to the initial disparity. Alternatively, the disparity remaining can be obtained by additionally modeling the intermediate confounder model. To be more specific, disparity remaining for intervention 1 is defined as

$$\zeta^1(1) = \sum_{s,m_1} E[Y|R=1,s,m_1,c]P(s|R=1,c)P(m_1|R=0,c) - E[Y|R=0,c], \quad (10)$$

where $s \in \mathcal{S}, m_1 \in \mathcal{M}_1$, and $c \in \mathcal{C}$. As before, $E[Y|R=0,c]$ can be rewritten as $\beta_0 + \beta_2 E[S|R=0,c] + \beta_3 E[M_1|R=0,c] + \beta_5 c$. Given this, the disparity remaining can be rewritten as

$$\begin{aligned}\zeta^1(1) =& \beta_1 + \beta_2\{E[S|R=1,c] - E[S|R=0,c]\} + \beta_4 E(M_1|R=0,c) \\ =& \beta_1 + \beta_2\kappa_1 + \beta_4\alpha_0,\end{aligned} \tag{11}$$

where $\kappa_1$ is the difference in $S$ between $R=1$ and $R=0$ given $C=c$. This completes the proof.

## Appendix B. A Proof for the Single-mediator-imputation Estimator (Conditional on Covariates)

Disparity reduction for intervention 1 is defined as

$$\delta^1(c) = E[Y|R=1,c] - \sum_{s,m_1} E[Y|R=1,s,m_1,c]P(s|R=1,c)P(m_1|R=0,c), \quad (12)$$

where $s \in \mathcal{S}, m_1 \in \mathcal{M}_1$, and $c \in \mathcal{C}$. Here, the latter quantity can be rewritten as

$$
\begin{aligned}
&= \sum_{S,R,m_1} I(R=1)E[Y|R,S,m_1,c]P(S|R,c)P(m_1|R=0,c) \\
&= \sum_{S,R,m_1} \frac{I(R=1)}{P(R=1|c)} E[Y|R,S,m_1,c]P(S|R,c)P(m_1|R=0,c)P(R|c) \\
&= \sum_{S,R,m_1} \frac{I(R=1)}{P(R=1|c)} E[Y|R,S,m_1,c]P(m_1|R=0,c)P(S,R|c) \\
&= E[\frac{I(R=1)}{P(R=1|c)} \sum_{m_1} E[Y|R,S,m_1,c]P(m_1|R=0,c)|c] \\
&= E[\frac{P(R=1|c)}{P(R=1|c)} \sum_{m_1} E[Y|R,S,m_1,c]P(m_1|R=0,c)|R=1,c] \\
&= E[\sum_{m_1} E[Y|R,S,m_1,c]P(m_1|R=0,c)|R=1,c].
\end{aligned}
\quad (13)
$$

The fourth equality is due to the law of iterated expectation with respect to $S$ and $R$ given $C = c$. The fifth equality is due to applying $P(A,B|c) = P(A|c)P(B|A,c)$. This completes the proof.

**Appendix C. A Proof for the Single-mediator-imputation Estimator (Marginal)**

Disparity reduction for intervention 1 is defined as

$$\delta^1(c) = E[Y|R=1,c] - \sum_{s,m_1} E[Y|R=1,s,m_1,c]P(s|R=1,c)P(m_1|R=0,c)P(c), \quad (14)$$

where $s \in \mathcal{S}, m_1 \in \mathcal{M}_1$, and $c \in \mathcal{C}$. Here, the latter quantity can be rewritten as

$$= \sum_{S,R,m_1,c} I(R=1)E[Y|R,S,m_1,c]P(S|R,c)P(m_1|R=0,c)P(c)$$

$$= \sum_{S,R,m_1,c} \frac{I(R=1)}{P(R=1|c)} E[Y|R,S,m_1,c]P(S|R,c)P(m_1|R=0,c)P(R|c)P(c)$$

$$= \sum_{S,R,m_1,c} \frac{I(R=1)}{P(R=1|c)} E[Y|R,S,m_1,c]P(m_1|R=0,c)P(S,R,c) \quad (15)$$

$$= E[\frac{I(R=1)}{P(R=1|c)} \sum_{m_1} E[Y|R,S,m_1,c]P(m_1|R=0,c)]$$

$$= E[\frac{P(R=1)}{P(R=1|c)} \sum_{m_1} E[Y|R,S,m_1,c]P(m_1|R=0,c)|R=1]$$

The fourth equality is due to the law of iterated expectation with respect to $S$, $R$, and $C=c$. The fifth equality is due to applying $P(A,B) = P(A)P(B|A)$. This completes the proof.